\documentclass[runningheads]{llncs}
\usepackage{amsmath}
\usepackage{graphicx}
\usepackage{algorithm}
\usepackage{algpseudocode}
\usepackage{comment}
\usepackage{placeins}

\begin{document}
\title{Self-organized MT Direction Maps Emerge from Spatiotemporal Contrastive Optimization\thanks{Preprint. Under review.}}
\titlerunning{MT Direction Maps Emerge from Spatiotemporal Contrastive Optimization}

\author{Zhaotian Gu\inst{1}\thanks{Corresponding authors.} \and  Molan Li\inst{1, 3} \and Jie Su\inst{3} \and Chang Liu\inst{1} \and Tianyi Qian\inst{3} \and Dahui Wang\inst{1, 2}\protect\footnotemark[2]}

\authorrunning{Z. Gu et al.}

\institute{School of System Science, Beijing Normal University, Beijing 100875, China \\
\email{\{zhaotiangu, moonlanlee, liu\_chang\}@mail.bnu.edu.cn} \and
State Key Laboratory of Cognitive Neuroscience and Learning, Beijing Normal University, Beijing 100875, China \\
\email{wangdh@bnu.edu.cn} \and
Qiyuan Laboratory, Beijing 100095, China \\ \email{\{sujie, qiantianyi\}@qiyuanlab.com}}

\maketitle

\begin{abstract}
The spatial and functional organization of the primate visual cortex is a fundamental problem in neuroscience. While recent computational frameworks like the Topographic Deep Artificial Neural Network (TDANN) have successfully modeled spatial organization in the ventral stream, the computational origins of the dorsal stream's distinct topographies, such as direction-selective maps in the middle temporal (MT) area, remain largely unresolved. In this work, we present a spatiotemporal TDANN to investigate whether MT topography is governed by the same universal principles. By training a 3D ResNet on naturalistic videos via a Momentum Contrast (MoCo) self-supervised paradigm alongside a biologically inspired spatial loss, we demonstrate the spontaneous emergence of brain-like direction maps and topological pinwheel structures. 
Crucially, we reveal that MT tuning properties, characterized by strong direction selectivity paired with a residual axial component, arise from a strict optimization trade-off between task-driven discriminative pressure and spatial regularization. The model's representations quantitatively match \textit{in vivo} macaque MT physiological baselines, including direction selectivity index, circular variance, and pinwheel density. These findings unify the computational origins of the ventral and dorsal streams, establishing a general mechanism for cortical self-organization.
\end{abstract}

\section{Introduction}
The spatial and functional organization of the visual cortex is a fundamental problem in neuroscience. Neural recordings have long revealed that the primary visual cortex (V1) exhibits distinct topographic structures for orientation selectivity, such as ``pinwheels'' across various species~\cite{kaschube2010universality}. 
Extensive computational models suggest these structures arise from a unifying hypothesis: the brain organizes its high-dimensional feature representations to optimize task-driven learning while satisfying stringent biophysical constraints (e.g., minimizing wiring length)~\cite{chklovskii2002wiringa,jacobs1992computationala}. Similarly, in the dorsal visual pathway, although the middle temporal (MT) area is highly specialized for complex spatiotemporal processing, it also harbors intricate topographic structures, including continuous direction-selective maps and pinwheel singularities~\cite{1994optical,diogo2003electrophysiological}. However, it remains largely unresolved whether the emergence of these distinct topographies in area MT is governed by the same universal self-organizing principles that shape V1 and the ventral stream.

While recent NeuroAI frameworks, such as the Topographic Deep Artificial Neural Network (TDANN)~\cite{margalit2024unifyinga}, have successfully modeled the emergence of spatial organization in the ventral stream (e.g., V1 orientation maps and VTC category-selective patches) using 2D static images, the computational origins of the dorsal stream's temporal dynamics remain unexplored from the same perspective.

In this work, we bridge this gap by extending the TDANN framework to the spatiotemporal domain. We hypothesize that the emergence of MT direction maps is driven by the exact same optimization principle: balancing unsupervised representation learning from dynamic visual experience with local spatial regularization. By employing a 3D ResNet-18~\cite{hara2017can} to process naturalistic videos and introducing a Momentum Contrast (MoCo)~\cite{he2020momentuma} self-supervised paradigm alongside a biologically-inspired spatial loss, we simulate the functional development of the dorsal visual pathway.

Our contributions are as follows:
\begin{enumerate}
    \item \textbf{Spatiotemporal Topographic Framework}: We implement the first spatiotemporal TDANN that successfully demonstrates the spontaneous emergence of brain-like direction-selective maps and topological ``pinwheel'' structures in an MT-like layer.
    \item \textbf{Mechanistic Insights into Cortical Tuning}: We reveal that MT tuning properties, characterized by strong direction selectivity paired with a residual axial component~\cite{rust2006how}, arise from a strict optimization trade-off between task-driven discriminative pressure and spatial regularization.
    \item \textbf{Physiological Alignment}: We demonstrate that the model's emergent representations exquisitely match \textit{in vivo} macaque MT physiological baselines, including direction selectivity index, circular variance, and macroscopic pinwheel density.
\end{enumerate}

\section{Related Work}
\paragraph{Traditional Models of Functional Organization.}
Early computational models of cortical maps relied on hand-crafted features or hard-coded connectivity rules~\cite{swindale1980model,linsker1986basic}.
While models like Kohonen's self-organizing maps successfully replicated orientation columns and ocular dominance patterns, they failed to explain how such structures naturally emerge from unsupervised sensory experience~\cite{kohonen1982selforganized,obermayer1990principle,yamins2016using,margalit2024unifyinga}. 
More recent connectivity-reconstruction models for the middle temporal (MT) area~\cite{koprinkova-hristova2019spike} provide biological realism but lack a unified task-driven learning objective. 
The key limitation across traditional approaches is the absence of a principle that simultaneously achieves both functional efficiency and spatial structure.

\paragraph{Topographic Deep Artificial Neural Networks (TDANNs).}
The TDANN framework~\cite{margalit2024unifyinga} represents a paradigm shift toward ``unifying function and spatial structure'' in a single learning objective. The key innovation is combining self-supervised learning (via SimCLR~\cite{chen2020simple}) with a spatial locality constraint that penalizes units far apart on a simulated 2D cortical sheet for having dissimilar response profiles. By optimizing a composite loss combining contrastive loss and balanced spatial loss, TDANN demonstrated that spontaneous emergence of organized maps and tuning properties in ventral temporal cortex, occurs without hand-coded structure.

However, the original TDANN was evaluated exclusively on 2D images, leaving fundamental questions unanswered: Do these self-organization principles generalize to spatiotemporal processing? Can they explain the dorsal stream's direction-selective organization in MT? Our work addresses this gap by extending TDANN's framework to the temporal domain.

\paragraph{Self-Supervised Learning.}
Self-supervised learning (SSL) has emerged as a principled approach for modeling biological visual systems. Unlike supervised learning with artificial labels, SSL learns representations from the statistical structure of natural stimuli itself, making it more biologically plausible~\cite{lecun201911}. Contrastive methods like SimCLR~\cite{chen2020simple}, BYOL~\cite{grill2020bootstrap}, and MoCo~\cite{he2020momentuma} achieve this by learning to map similar views of the same instance to nearby points in representation space, with negative samples providing contrast.
We leverage this property to train our spatiotemporal TDANN, hypothesizing that combining motion-sensitive SSL with spatial locality constraints will drive the spontaneous emergence of direction-selective maps.

\section{Proposed Method}
\begin{figure}[t]
\centering 
\includegraphics[width=0.96\linewidth, trim=0 5pt 0 5pt, clip]{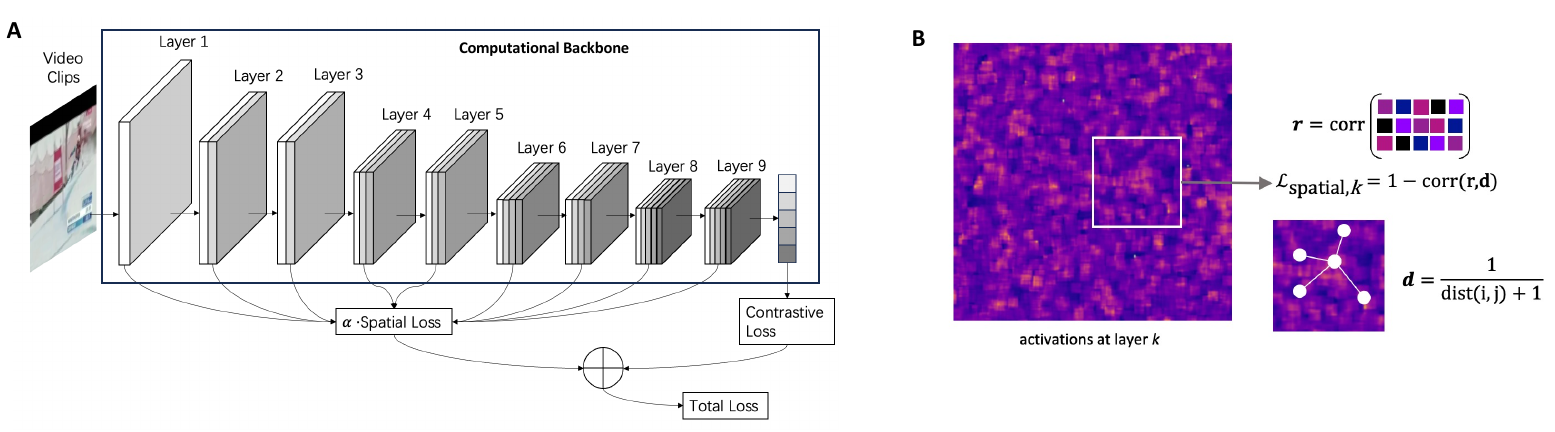} 
\caption{Spatiotemporal TDANN overview. \textbf{A} 3D ResNet-18 backbone with MoCo and spatial losses. \textbf{B} Simulated cortical sheet and spatial loss that promotes similar responses among nearby units.} 
\label{fig:1} 
\end{figure}

\paragraph{Backbone Architecture and Biological Mapping.}
We use a 3D ResNet-18 architecture as the computational backbone (Fig.~\ref{fig:1})~\cite{hara2017can} to integrate spatiotemporal signals, mirroring multi-scale biological motion perception through the hierarchially growing temporal receptive fields. 
We establish a formal mapping between network layers and biological brain regions based on physiological parameters and anatomical principles~\cite{vanni2020anatomy,dacey1992dendritic}, as detailed in Table \ref{tab:layer_mapping}. Layer 7 is designated as the ``MT-like'' layer because its position in the computational hierarchy matches biological MT's role between early visual areas (V1, V2) and higher motor-planning regions (LIP).

For 3D CNNs, activations contain both spatial and temporal information. When mapping network units to cortical locations, we consider only spatial information and average over time to simulate neural firing rates, yielding the spatial topography for analysis.

\begin{table}[t] 
\centering
\small
\caption{Layer Mapping and Physiological Parameters (adapted from~\cite{vanni2020anatomy,dacey1992dendritic})}
\begin{tabular}{|l|l|l|l|l|}
\hline
Model Layer & Units & Cortical Area ($mm^2$) & Neighborhood ($mm$) & Biological Area \\
\hline
Layer 2-3 & 200,704 & 5.7 & 0.047 & Retina \\
Layer 4-5 & 100,352 & 1,180 & 2.7 & V1 \\
Layer 6 & 50,176 & 940 & 4.2 & V2 \\
\textbf{Layer 7} & \textbf{50,176} & \textbf{50} & \textbf{2.1} & \textbf{MT} \\
Layer 8-9 & 25,088 & 56 & 2.7 & LIP \\
\hline
\end{tabular}
\label{tab:layer_mapping}
\end{table}

The cortical area values match empirical measurements of primate anatomy, and the neighborhood column specifies local connectivity radius grounded in
anatomical literature, defining the scale at which spatial constraints enforce local functional similarity. 

\paragraph{Self-Supervised Paradigm: Momentum Contrast (MoCo).}
Following the TDANN hypothesis that neural systems optimize for ecologically relevant behavior~\cite{margalit2024unifyinga}, we employ MoCo for instance discrimination on the UCF101 dataset~\cite{soomro2012ucf101}. 

Given a query $q$ and a positive key $k^+$ (two temporally-augmented clips from the same video), the contrastive loss is:
\begin{equation}
\mathcal{L}_{\text{contrast}} = -\log \frac{\exp(q \cdot k^+ / \tau)}{\exp(q \cdot k^+ / \tau) + \sum_{i=1}^K \exp(q \cdot k_i^- / \tau)}
\end{equation}
where $\tau$ is the temperature parameter controlling the softness of the distribution, and $K$ is the number of negative samples in the momentum queue. This objective forces the encoder to extract invariant temporal features while ignoring frame-specific spatial details. The result is a representation optimized for capturing motion statistics that naturally drive direction selectivity in the dorsal stream.

\paragraph{Spatial Locality Constraint.}
To simulate cortical biophysical efficiency, where nearby neurons share similar functions to minimize wiring costs~\cite{hubel1962receptive,jacobs1992computationala,chklovskii2002wiringa}, we introduce a spatial loss $\mathcal{L}_{\text{spatial},k}$ for each layer $k$:
\begin{equation}
\mathcal{L}_{\text{spatial},k} = \frac{1 - \text{corr}(\mathbf{r}_k, \mathbf{D}_k)}{2}
\end{equation}
where $\mathbf{r}_k$ represents pairwise unit activation correlations and $\mathbf{D}_k$ is the inverse physical distance vector ($D_i = \frac{1}{d_i + 1}$) on the simulated 2D cortical sheet (Fig.~\ref{fig:1}\textbf{B}). The total objective balances task performance with spatial organization:
\begin{equation}
\mathcal{L}_{\text{total}} = \mathcal{L}_{\text{contrast}} + \alpha \sum_{k} \mathcal{L}_{\text{spatial},k}
\end{equation}

Spatial constraint $\alpha$ controls the trade-off between representation learning and biophysical efficiency. This competition forces informative representations to organize locally, driving the emergence of functional maps.

\paragraph{Multi-Step Progressive Training Strategy.}

Optimizing spatial structure in weight-sharing CNNs presents a fundamental conflict: gradient updates to shared filters affect all spatial units simultaneously, making direct spatial optimization highly unstable. To decouple representation learning from spatial organization while maintaining their cooperative synergy, we employ a six-step progressive strategy:

\begin{enumerate}
    \item \textbf{Representation Pre-training:} Train the 3D ResNet-18 using only the contrastive loss ($\mathcal{L}_{\text{contrast}}$) on UCF101 to establish robust motion features.
    \item \textbf{Initial Position Initialization:} Initialize unit positions based on biological feedforward hierarchy to preserve coarse retinotopy (Appendix~\ref{appendix:position_initialization}).
    \item \textbf{Iterative Position Pre-optimization:} Iteratively rearrange unit positions on the simulated cortical sheet so units with correlated motion responses are placed closer together (Appendix~\ref{appendix:spatial_rearrangement}). 
    \item \textbf{Position Freezing:} Permanently freeze all unit positions to preserve topographic structure and eliminate the weight-sharing conflict.
    \item \textbf{Weight Re-initialization:} Randomly reset all network weights. This critical step decouples the model from the weight history of Step 1.
    \item \textbf{Joint Training:} Re-train the network from scratch with frozen positions using the combined loss ($\mathcal{L}_{\text{total}}$). 
\end{enumerate}

This pipeline models a biologically plausible developmental process. The position pre-optimization serves as an experience-independent structural scaffold, mimicking coarse topographic seeding by pre-natal retinal waves~\cite{ge2021retinal}. Subsequently, joint training (Steps 5-6) simulates experience-dependent refinement: starting from random weights, the network aligns its task-driven representation learning with this fixed spatial prior, driving the emergence of biologically constrained brain-like representations.

\section{Layer Response and Statistical Alignment}
\begin{figure}[t]
\centering 
\includegraphics[width=0.96\linewidth]{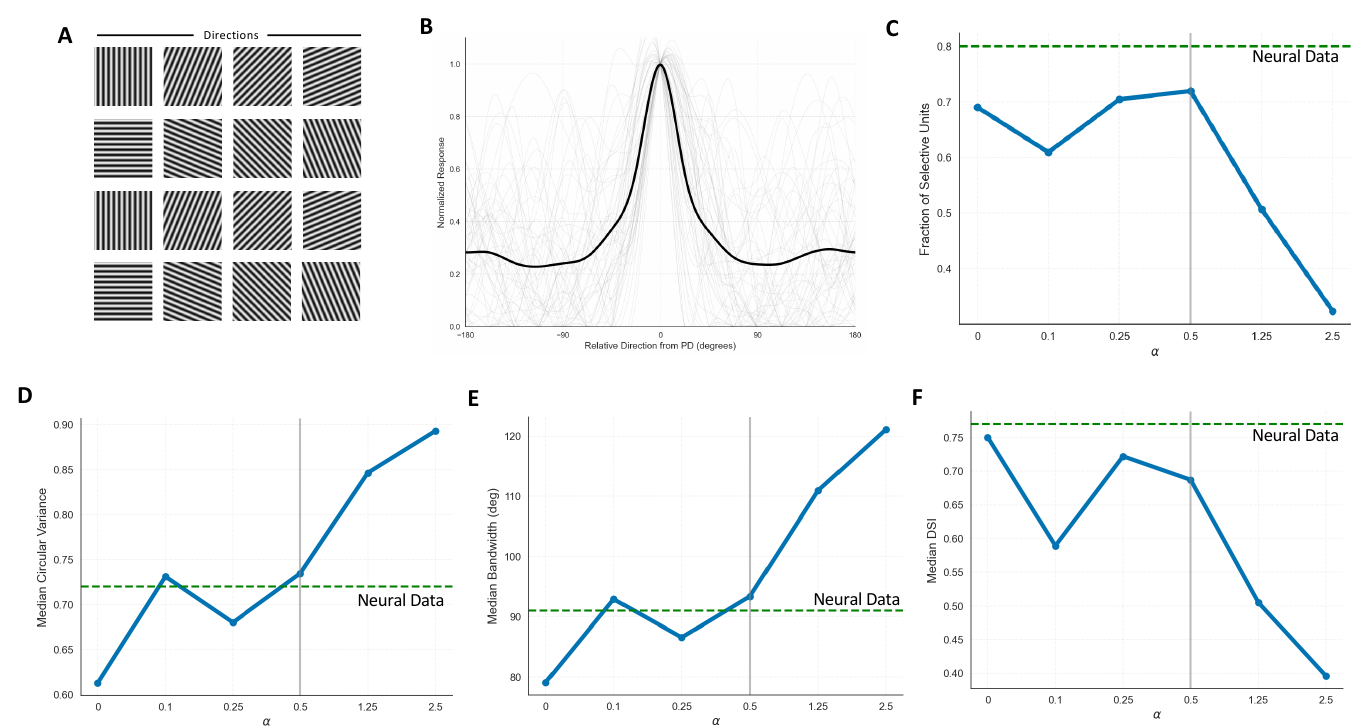} 
\caption{Biological validation and parameter sensitivity of emergent tuning properties in the MT-like layer.
\textbf{A} 16 drifting-grating directions used for probing.
\textbf{B} Population tuning curve (PTC) at $\alpha=0.5$.
\textbf{C-F} Sensitivity to $\alpha$: selective-unit fraction (\textbf{C}), median bandwidth(\textbf{D}), median CV(\textbf{E}), and median DSI(\textbf{F}); green dashed lines indicate macaque MT baselines~\cite{nakhla2021neural,albright1984direction,britten1992analysis}. Gray line: $\alpha=0.5$.}
\label{fig:2} 
\end{figure}

To assess the functional organization of the model's MT-like layer, we stimulated the network with drifting gratings at 16 directions (Fig.~\ref{fig:2}\textbf{A}, Appendix~\ref{appendix:drifting_grating_stimuli}). We evaluated the emergent representations through the Direction Selectivity Index (DSI) and the Population Tuning Curve (PTC).

\paragraph{Functional Classification of the MT-like Layer.}
Following established physiological criteria~\cite{albright1984direction,maunsell1983functional}, we categorized the units in the MT-like layer according to their DSI. Under the spatial constraint of $\alpha = 0.5$, the layer exhibited a high degree of functional specialization: approximately 72\% of the units demonstrated significant direction selectivity ($\text{DSI} > 0.5$, Fig.~\ref{fig:2}\textbf{C}). This population census aligns with macaque MT data, where $\sim 80\%$ of neurons exhibit directional preferences~\cite{nakhla2021neural,albright1984direction,britten1992analysis}.
The median DSI of the MT-like layer was approximately 0.68 (Fig.~\ref{fig:2}\textbf{C}), in close to the physiological baseline of $\sim 0.77$ reported in macaque MT~\cite{nakhla2021neural,albright1984direction}.

\paragraph{PTC Morphology and Statistical Matching.}
The emergent tuning profiles were visualized via the PTC (Fig.~\ref{fig:2}\textbf{B}). The MT-like layer exhibits a well-defined, sharp primary peak centered at the preferred direction, indicating a high degree of directional discriminability. The response is effectively suppressed at the sideflaps, while a characteristic secondary rebound (approx. 0.3 magnitude) peak is observed at the opposite direction,
suggesting that while units remain predominantly direction-selective, they retain a residual axial bias. This bimodal profile reflects the inherent structure of spatiotemporal filters in our model, where the pressure for direction discrimination (suppressing the opposite direction) is partially moderated by the spatial constraint to maintain topographic continuity~\cite{adelson1985spatiotemporal}.

Quantitatively, the MT-like population exhibits a median Circular Variance (CV) of approximately 0.73 (Fig.~\ref{fig:2}\textbf{E}), closely matching the physiological baseline of $\sim 0.72$ reported in macaque MT~\cite{nakhla2021neural,albright1984direction}. Using the wrapped normal approximation ($\sigma = \sqrt{-2\ln(1-\text{CV})}$), this CV translates to a median statistical bandwidth of $93.36^\circ$ (Fig.~\ref{fig:2}\textbf{D}). This statistical profile provides an exquisite match to the \textit{in vivo} MT baseline of $\sim91^\circ$ reported before~\cite{nakhla2021neural,albright1984direction}.

\section{Optimization Trade-offs: Discriminative Pressure vs. Spatial Regularization}
\begin{figure}[t]
\centering 
\includegraphics[width=0.96\linewidth, trim=0 5pt 0 2pt, clip]{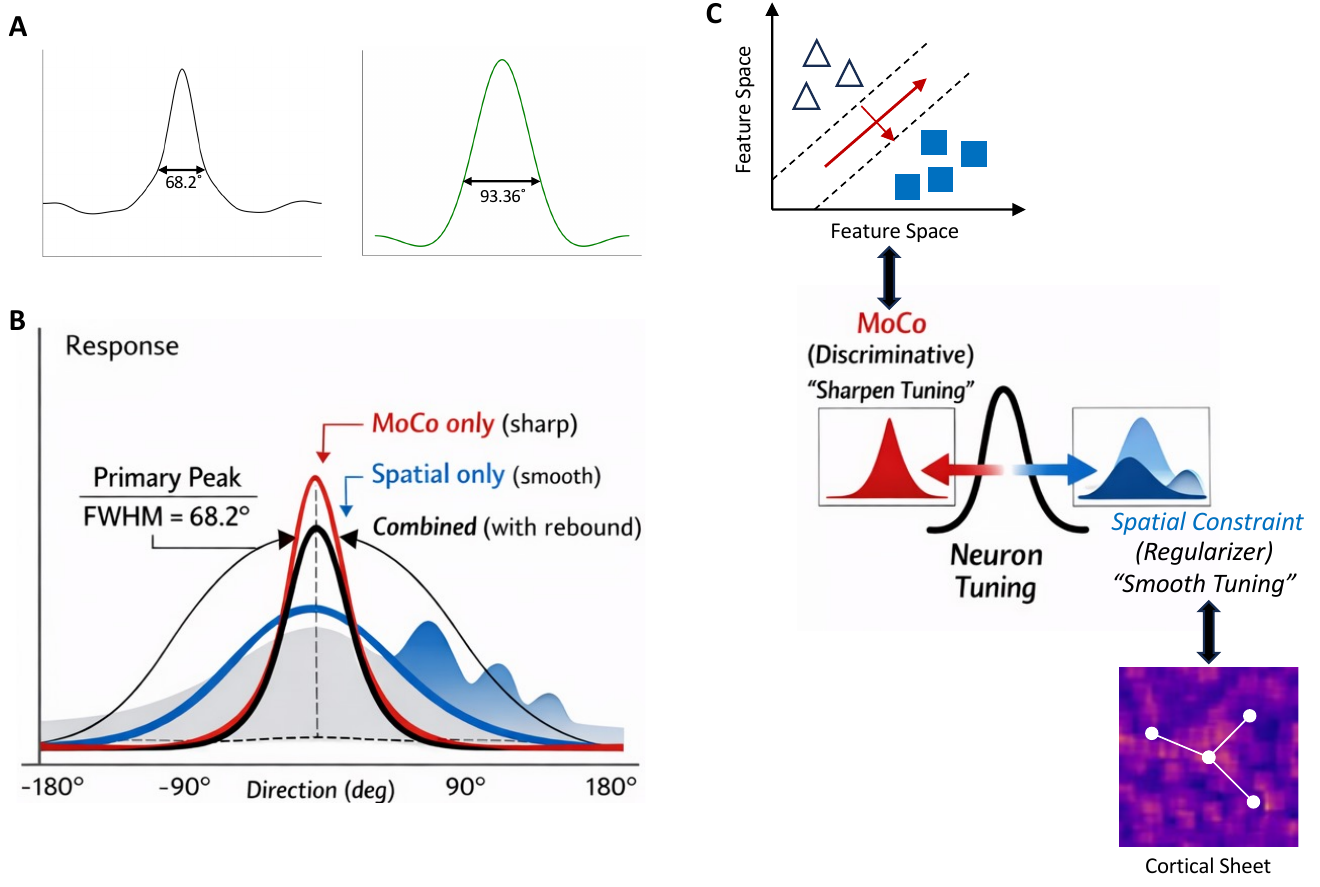} 
\caption{Mechanistic origins of emergent tuning properties through optimization trade-offs.
\textbf{A} FWHM and CV-derived bandwidth diverge in the optimal model.
\textbf{B} Decomposition of tuning under contrastive-only, spatial-only, and joint objectives.
\textbf{C} Schematic of competing dynamics between discriminative pressure and spatial smoothness.}  
\label{fig:3} 
\end{figure}

A distinctive feature of our optimal model ($\alpha=0.5$) is the divergence between the direct geometric measurement of the primary peak ($\text{FWHM} = 68.2^\circ$) and the CV-derived statistical bandwidth ($\sim93^\circ$, Fig.~\ref{fig:3}\textbf{A}). We interpret this phenomenon as a consequence of the direct competition between the two training objectives, which shapes a tuning profile characterized by strong direction selectivity alongside a residual axial bias (Fig.~\ref{fig:appendix:ptc_alpha} in Appendix~\ref{appendix:ptc_alpha}).

\paragraph{Discriminative Pressure.}
In the absence of spatial constraints ($\alpha=0$), the Momentum Contrast (MoCo) objective dominates. Because contrastive learning optimizes for instance discrimination~\cite{wu2018unsuperviseda,he2020momentuma}, it drives the network to develop sharp direction selectivity to separate spatiotemporal inputs~\cite{wang2020understanding,wen2021understanding,zhuang2021unsupervised}, producing a baseline FWHM of $61.5^\circ$. Interestingly, as mild spatial pressure is introduced ($\alpha \in [0.1, 0.25]$), the network faces an optimization conflict. To preserve latent discriminability while accommodating spatial sharing, it paradoxically sharpens the primary peak (dropping to $49.7^\circ$). This sharpening is accompanied by an emerging $180^\circ$ residual component, reflecting the inherent axial symmetry of spatiotemporal energy filters that the network leverages to maintain representational stability under constraint.

\paragraph{Spatial Regularization.}
The spatial loss ($\mathcal{L}_\text{spatial}$) acts as a global regularizer that penalizes dissimilar responses among neighboring units. While MT neurons are predominantly direction-selective, 3D convolutional kernels are mathematically rooted in spatiotemporal orientations~\cite{adelson1985spatiotemporal,hara2017can}. To satisfy local smoothness across the cortical sheet~\cite{margalit2024unifyinga,durbin1990dimension}, the network `tolerates' a residual bimodal structure rather than pursuing absolute unidirectional suppression (Fig.~\ref{fig:3}\textbf{B\&C}). At the optimal spatial constraint ($\alpha=0.5$), the primary peak relaxes to $68.2^\circ$, and this bimodal balance stabilizes. Because Circular Variance is sensitive to the entire $360^\circ$ profile, the presence of this residual axial component inflates the statistical bandwidth to $\sim93^\circ$, providing a normative explanation for the global statistical profile observed in macaque MT~\cite{albright1984direction,nakhla2021neural}.

\section{Emergence of ``Pinwheel'' Structure in Direction Map}
\begin{figure}[t]
\centering 
\includegraphics[width=0.96\linewidth, trim=0 5pt 0 5pt, clip]{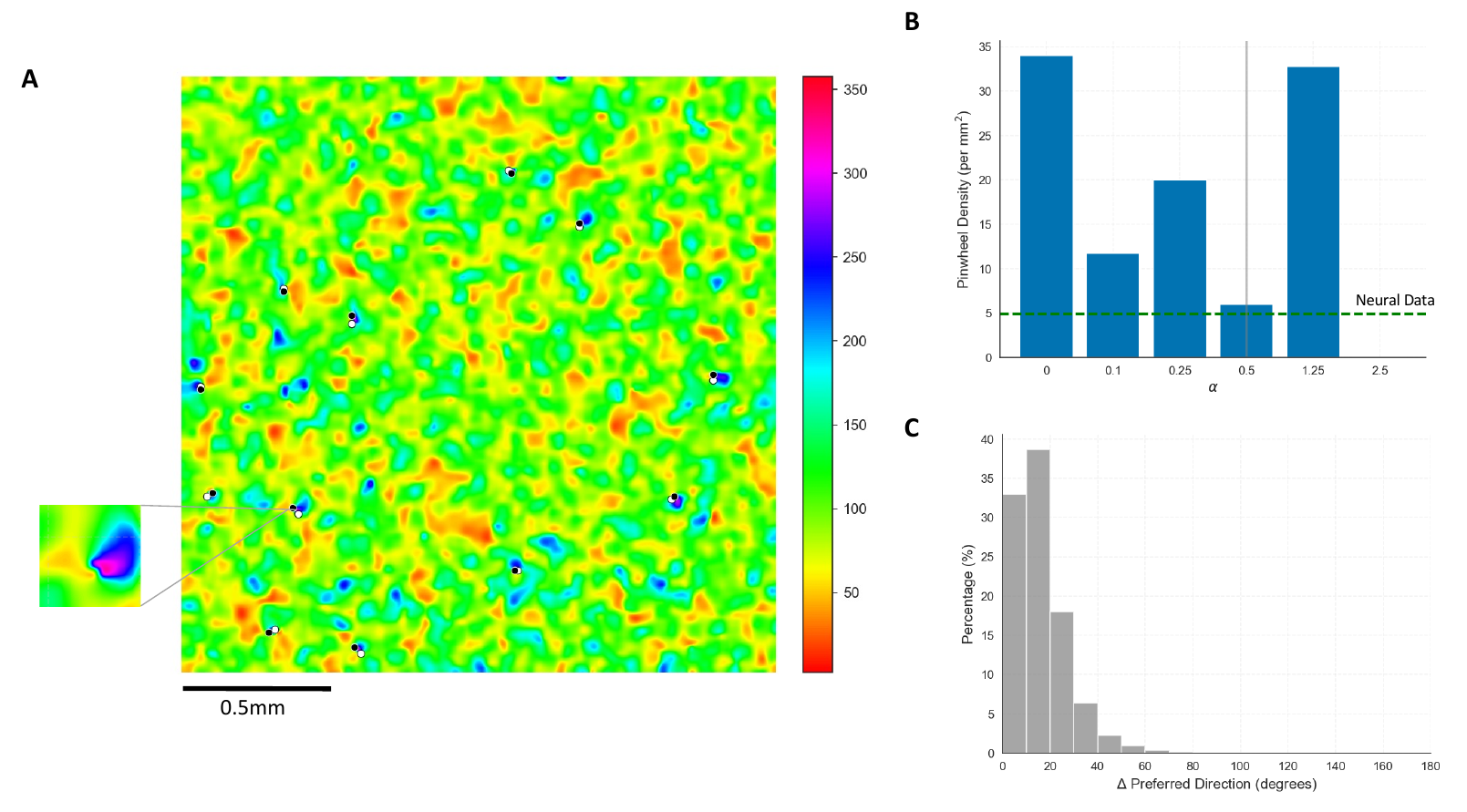} 
\caption{Spontaneous emergence and quantitative analysis of MT-like pinwheel structures.
\textbf{A} Direction map at $\alpha=0.5$ with positive (white dot) and negative (black dot) pinwheel centers.
\textbf{B} Pinwheel density versus $\alpha$; green dashed line marks the macaque MT baseline ($\sim 4.9\,\text{mm}^{-2}$) estimated from~\cite{1994optical,kaschube2010universality}. Gray line: $\alpha=0.5$.
\textbf{C} Histogram of preferred direction differences ($\Delta$) between adjacent units at $\alpha=0.5$.}
\label{fig:4} 
\end{figure}

To analyze the topographic organization of the MT-like layer, we simulate a physiological recording paradigm: unit responses are aggregated into a spatial grid and processed with Gaussian smoothing to mimic the spatial resolution and point-spread function of optical imaging (see Appendix~\ref{appendix:neural_readout} for readout details). This procedure reveals the spontaneous emergence of singularities, 
the defining topographic hallmark of primate MT is the ``pinwheel'' structure~\cite{diogo2003electrophysiological,1994optical}. As visualized in a representative $2 \text{mm} \times 2 \text{mm}$ sub-region of our MT-like layer (Fig.~\ref{fig:4}\textbf{A}, $\alpha=0.5$), the network spontaneously develops continuous iso-direction patches interspersed with positive (+1) and negative (-1) topological singularities (see Appendix~\ref{appendix:singularity_detection} for detection details). These singularities frequently organize into dipole pairs, a characteristic low-energy topological state that maintains map stability~\cite{kosterlitz1973ordering,ribot2016pinwheeldipole,kaschube2010universality}.

\paragraph{Optimization-Driven Topography.}
Mirroring the trade-offs observed in neural tuning, map topography might also be governed by the competition between discriminative pressure and spatial regularization (Fig.~\ref{fig:4}\textbf{B}). Under pure discriminative pressure ($\alpha=0$), independent unit optimization generates a hyper-fragmented map with redundant topological defects(Fig.~\ref{fig:appendix:maps_alpha} in Appendix~\ref{appendix:maps_alpha}). Introducing optimal spatial regularization ($\alpha=0.5$) forces local representational sharing, annihilating redundant singularities to form smooth slabs. In our analyzed $4 \text{mm}^2$ patch, this balance yields 24 pinwheel centers (a density of $6 \text{mm}^{-2}$), dropping the pinwheel count to a local minimum that aligns with primate physiological baselines estimated from~\cite{1994optical,kaschube2010universality} (Appendix~\ref{appendix:estimation_pinwheel_density}).
Conversely, the intense objective conflict at $\alpha=1.25$ identically manifests as spatial instability, shattering the map into a high-density state before collapsing into uniformity at $\alpha=2.5$ (Fig.~\ref{fig:appendix:maps_alpha} in Appendix~\ref{appendix:maps_alpha}).

\paragraph{Functional Hubs and Smoothness.}
This optimal topography is quantitatively verified by local direction gradients within the sub-region (Fig.~\ref{fig:4}\textbf{C}). Angular differences are heavily concentrated below $20^\circ$, confirming robust local smoothness. Functionally, these pinwheels act as indispensable topological hubs that guarantee complete $360^\circ$ directional coverage within any localized cortical neighborhood~\cite{swindale1980model,durbin1990dimension,albright1984direction}. Thus, pinwheels emerge as the ultimate geometric solution balancing biophysical wiring efficiency and complete task-driven feature representation.

\section{Discussion and Conclusion}
\paragraph{General Principles of Cortical Organization.}
Our results extend the TDANN framework to the spatiotemporal domain, unifying the computational origins of the ventral and dorsal visual streams. We demonstrate that topographic organization, specifically the emergence of MT direction maps, arises fundamentally from a strict optimization trade-off. The competition between task-driven discriminative pressure and biophysical spatial constraints serves as a universal mechanism shaping both individual neural tuning and global map topography.

\paragraph{Mechanistic Insights into the Dorsal Stream.}
By dissecting this trade-off, our model provides a normative explanation for key physiological hallmarks of the primate MT. 
Functionally, the network resolves the conflict between latent discriminability and spatial sharing by retaining an axial bimodal component rather than pursuing absolute unidirectional suppression, and this trade-off explains the divergence between geometric FWHM and statistical bandwidth.
Topographically, ``pinwheel'' emerge not as developmental artifacts, but as indispensable topological hubs, i.e. the optimal geometric solution to guarantee complete $360^\circ$ directional coverage within localized cortical patches under stringent wiring constraints.

\paragraph{Limitations and Future Work.}
While our spatiotemporal TDANN captures topographic characteristics, it lacks the temporal recurrence and top-down feedback prevalent in biological brains. Future iterations should incorporate Convolutional Recurrent Neural Networks (ConvRNNs)~\cite{nayebi2018taskdriven} or Spiking Neural Networks (SNNs), and explore biologically realistic learning rules (e.g., Hebbian plasticity~\cite{shaw1986donald,oja1982simplified} or predictive coding~\cite{rao1999predictive}). Additionally, rigorous quantitative comparisons with electrophysiological data during active behavioral tasks (e.g., motion tracking) will further validate the model.

\paragraph{Conclusion.}
We demonstrate that the functional organization of the primate MT area spontaneously emerges from the synergy of self-supervised motion representation and spatial locality constraints. This work establishes a neurobiologically plausible framework for cortical self-organization, offering a unified principle that bridges machine learning architecture design and primate visual system development.

\bibliographystyle{splncs04}
\bibliography{refs.bib}

\appendix

\section{Appendix: Position Initialization} \label{appendix:position_initialization}

The position initialization algorithm establishes biologically-plausible initial positions by: (1) computing receptive field centers and radii from layer parameters (Table~\ref{tab:layer_mapping}); (2) generating anchor points and applying random/grid-based offsets, Gaussian jitter, and scaling-translation to produce the initial position set; (3) randomly sampling neighboring windows and recording unit indices within each window. Neighborhoods are trimmed to equalize neighbor counts, producing a rectangular neighbor index matrix. The algorithm outputs: unit positions, neighbor index matrix, and neighborhood radius, which are subsequently used in Steps 3 and 6 for spatial loss computation.

\section{Appendix: Spatial Rearrangement} \label{appendix:spatial_rearrangement}
To optimize spatial structure despite the constraints of weight-sharing CNNs, we employ an iterative rearrangement process inspired by prenatal retinal waves~\cite{ge2021retinal}. The simulated cortical sheet is partitioned into local neighborhoods where unit positions are stochastically swapped. A swap is retained only if it reduces the spatial loss ($\mathcal{L}_{\text{spatial}}$) by decreasing the physical distance between units with highly correlated activation features. We performed 10,000 total iterations across 20 neighborhoods, with the maximum unit count and neighborhood widths constrained by the biological parameters in Table~\ref{tab:layer_mapping} to ensure hierarchical consistency across visual areas.

\section{Appendix: Training Details} \label{appendix:training_details}
This model was trained on four NVIDIA A30 GPUs. Hyperparameters were selected based on best practices for video self-supervised learning using MoCo\cite{pan2021videomoco,qian2021spatiotemporal}. For each training stage, the model was trained for 200 epochs. For the first 10 epochs, the learning rate was linearly increased from 0.001 to a base learning rate of 0.03, and then gradually decreased to 0 using cosine annealing in subsequent training. Each batch contained 64 samples, and gradients from four batches were accumulated to update the model parameters. For each sample, data augmentation was performed, randomly cropping its size to $224 \times 224$ and extracting 16 frames. The momentum coefficient $m$ in MoCo was set to 0.999, and the temperature coefficient $\tau$ was set to 0.1.

\section{Appendix: Drifting Grating Stimuli} \label{appendix:drifting_grating_stimuli}
The drifting grating stimuli used to evaluate the model's tuning properties were generated according to
\begin{equation}
I(x, y, t) = \frac{C \cdot \sin(2\pi f (x \cos \theta + y \sin \theta - vt)) + B}{2}
\end{equation}
where $C$ is the contrast, $f$ is the spatial frequency, $\theta$ is the direction of motion, $v$ is the speed of the grating, and $B$ is the mean luminance.
In our experiments, we set $C=1$, $f=0.05$ cycles/pixel, $v=4$ pixel/frame, and $B=1$. 
The model was stimulated with drifting gratings at 16 different directions ($0^\circ, 22.5^\circ, 45^\circ, \cdots, 337.5^\circ$) to probe tuning properties and assess the direction selectivity of the emergent representations in the MT-like layer.

\section{Appendix: Neural Readout Processing} \label{appendix:neural_readout}
To simulate physiological recordings, unit responses were aggregated into a $700 \times 700$ grid ($10 \mu \text{m}$ per bin) covering the $50 \text{mm}^2$ MT layer. We applied a 2D Gaussian filter ($\sigma=10\mu\text{m}$) to eliminate sampling noise while preserving fine-grained topographic structures. This readout scale reflects the local integration of neural signals, from which direction maps of the MT-like layer were derived.

\section{Appendix: Detection of Topological Singularities} \label{appendix:singularity_detection}
To identify pinwheel centers, we computed the discrete topological charge $q$ for each $2 \times 2$ grid neighborhood by summing the wrapped phase differences along its closed contour: $q = \frac{1}{2\pi} \sum_{k=1}^{4} \mathcal{W}(\phi_{k+1} - \phi_k)$, where $\mathcal{W}$ wraps angle differences to $[-\pi, \pi]$. Grid locations yielding $q \approx +1$ or $-1$ denote positive (clockwise, white dots) and negative (counter-clockwise, black dots) topological singularities, respectively.

\section{Appendix: Estimation of Pinwheel Density} \label{appendix:estimation_pinwheel_density}
Following the universality principle where dimensionless pinwheel density $\rho \approx \pi$~\cite{kaschube2010universality}, we estimated the biological baseline for area MT. Using the empirical spatial wavelength $\Lambda \approx 0.8\text{ mm}$ measured in owl monkeys via optical imaging~\cite{1994optical}, the absolute density is derived as $\rho_{abs} = \pi / \Lambda^2 \approx 4.9\text{ centers/mm}^2$. This value serves as the structural constraint for our MT direction map model, representing a coarser functional grain compared to the V1 baseline ($\rho_{abs} \approx 12.5\text{ mm}^{-2}$).

\section{Appendix: PTC across Varying $\alpha$} \label{appendix:ptc_alpha}
\begin{figure}[H]
\centering 
\includegraphics[scale=0.4]{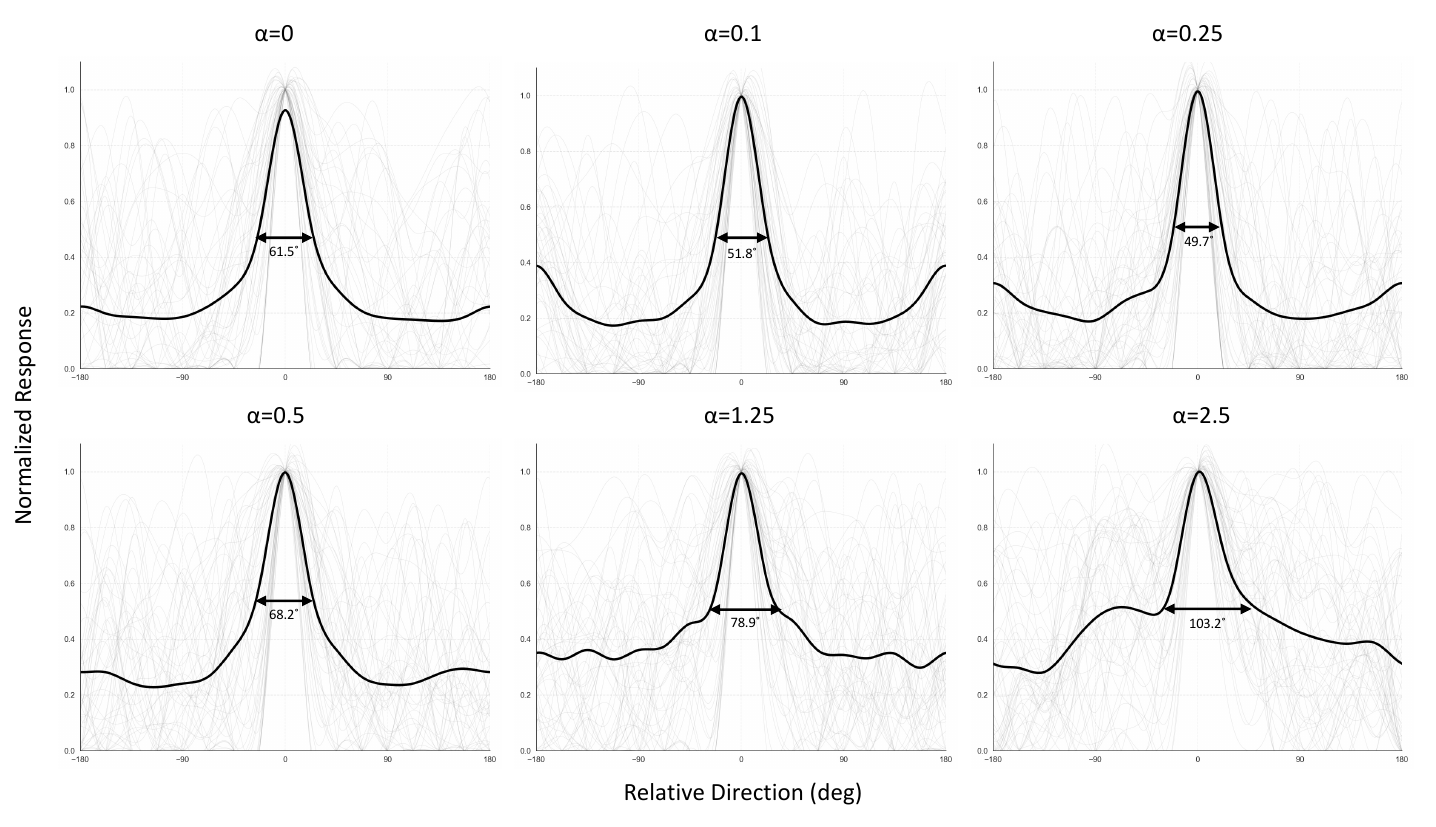} 
\caption{Population Tuning Curves (PTCs) across spatial constraints ($\alpha$), with annotated primary-peak FWHM values.}
\label{fig:appendix:ptc_alpha}
\end{figure}

\section{Appendix: Direction Maps across Varying $\alpha$} \label{appendix:maps_alpha}
\begin{figure}[H]
\centering 
\includegraphics[width=0.92\linewidth]{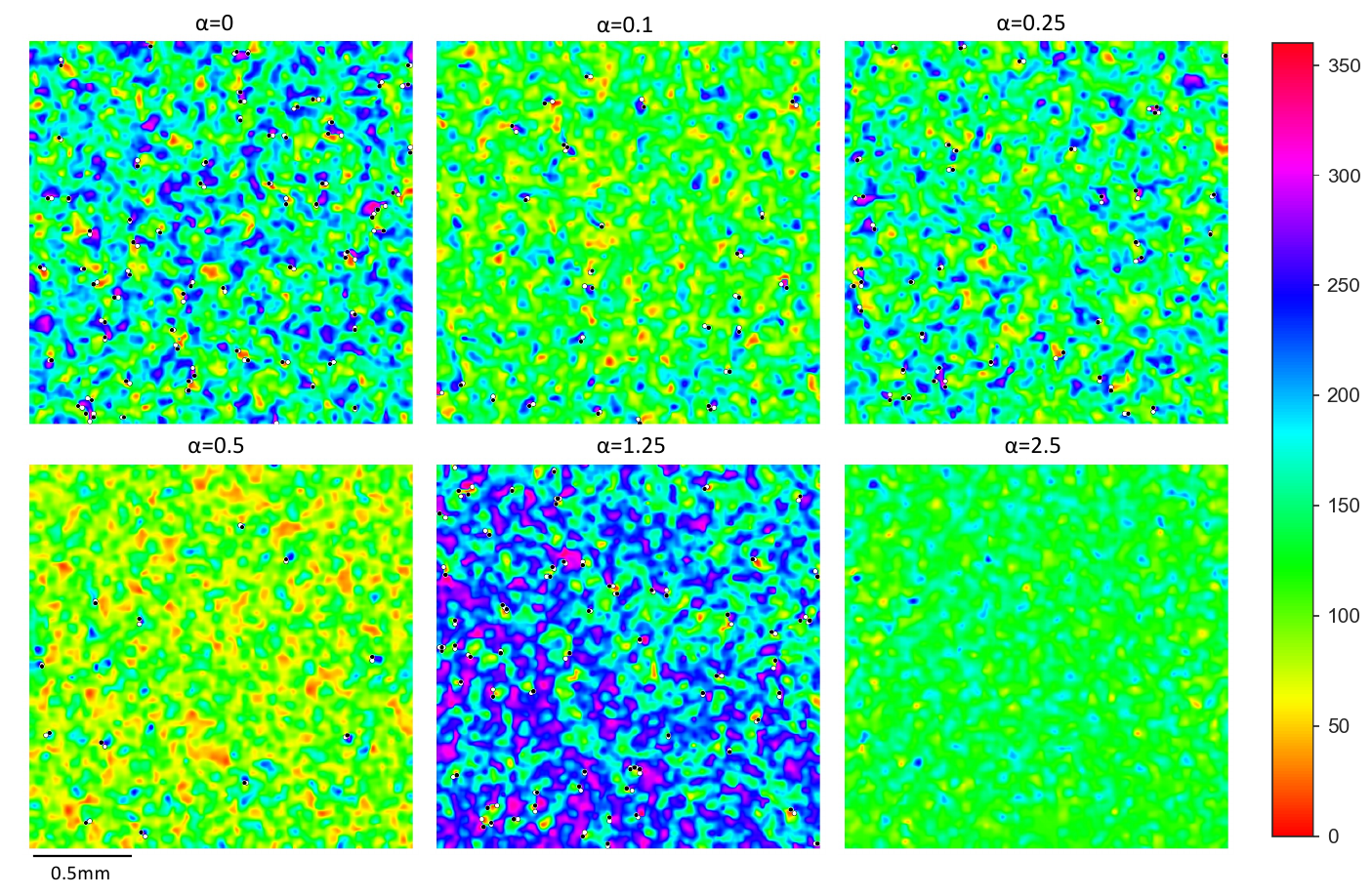}
\caption{Topological evolution of MT direction maps across $\alpha$ in a $2\,\text{mm} \times 2\,\text{mm}$ MT-like sub-region. White/black dots denote positive/negative pinwheel charges.}
\label{fig:appendix:maps_alpha}
\end{figure}

\end{document}